\documentclass[draft]{agujournal2019}
\usepackage{url} 
\usepackage[inline]{trackchanges} 
\usepackage{soul}
\draftfalse

\journalname{Journal of Geophysical Research: Solid Earth}
\journalname{}

\begin{document}

%
%


\title{VORA: Rapid Association of Earthquake Phases from Local to Global}

%
%




\authors{Junhao Song\affil{1,2}, Weiqiang Zhu\affil{1,2}, Haoyu Wang\affil{1,2}, Utpal Kumar\affil{1,2}, Taka'aki Taira\affil{1,2}, Richard M Allen\affil{1,2}}


\affiliation{1}{Department of Earth \& Planetary Science, University of California, Berkeley}
\affiliation{2}{Berkeley Seismology Laboratory, University of California, Berkeley}




\correspondingauthor{Junhao Song}{sjh2019@berkeley.edu}



\begin{keypoints}
\item We propose an earthquake phase association method (VORA) based on unsupervised clustering
\item We apply Voronoi tessellation to define station neighborhoods that adapt to changing network geometries across different regions and scales
\item We demonstrate that VORA achieves fast runtimes and scales from local networks to a globally distributed station set
\end{keypoints}

%
%

%
%


\begin{abstract}
Earthquake phase association, which groups seismic phase arrivals into common origins, is a key step towards more complete and reliable seismicity catalogs. It has become a challenging task because of the massive phase datasets produced from dense seismic networks and advanced phase picking methods. Here we present VORA (\textbf{V}oronoi tessellation- and \textbf{O}rigin-time-based \textbf{R}apid \textbf{A}ssociator), an efficient and scalable earthquake phase associator that treats association as an unsupervised spatio-temporal clustering problem. Specifically, VORA depends on two primary constraints: the estimated earthquake origin time (temporal) and seismic station adjacency (spatial). Recent advances in deep learning models have enabled detection of S- and P-phases with similar effectiveness, making it straightforward to estimate corresponding earthquake origin times from candidate phase pairs given a prescribed range of velocity ratios or directly from raw waveforms. We then leverage the Voronoi diagram to define each station's neighbors, cluster the origin times across neighboring stations, and apply an optional sub-clustering step to separate overlapping events. 
Benchmarks on synthetic and real datasets show that VORA achieves the fastest runtime and maintains robust performance (high recall and precision) even under intense seismicity. Applying it to a two-week global-scale pick dataset covering the 2019 Ridgecrest earthquake sequence further demonstrates that the same framework scales from local networks to a global station set. VORA requires no training and generalizes across local-to-global regions, varying velocity models, and evolving network geometries, helping to meet the growing demands of expanding seismic networks and the increasing volume of automated phase picks.
\end{abstract}

\section*{Plain Language Summary}

Earthquake signals are usually captured by surrounding seismic stations, where their P- and S-phase arrivals can be identified. Dense seismic networks and sensitive seismic phase picking techniques produce massive amounts of phase arrivals, especially during an intense earthquake sequence, making it more difficult to group phase arrivals into common earthquake origins. Because the ratio of P- to S-wave speed equals ratio of S- to P-wave travel time, we can estimate an earthquake's origin time based on its P and S phase arrival pairs, or directly from raw waveforms using machine learning methods. The origin times estimated for a single earthquake should be close in time and come from stations near its location. Therefore, we can perform spatio-temporal clustering of these hypothetical origin times from different stations to detect earthquakes. Based on this idea, we develop a new method named VORA and demonstrate its performance on both synthetic and real phase datasets from local to global spatial scales. The results show that VORA achieves the fastest runtime and maintains robust performance under various scenarios. 

%
%

%


%
%
%
%

\section{Introduction}

Earthquake monitoring plays an important role in mitigating seismic hazards and advancing seismological research. Its workflow generally comprises two sequential steps: detection and subsequent characterization. Detection involves distinguishing earthquake signals from background ambient noise through phase picking, followed by phase association to link discrete triggers into a common earthquake origin. The characterization stage further determines the source parameters, including time, location, magnitude, and focal mechanism, along with the resulting intensity of ground shaking. 
Similar processing sequences can be optimized for ultra-rapid earthquake early warning characterization of an event within a fraction of a second, through more traditional detection that uses minutes of data, to post-event research focused on developing the most complete catalog for a sequence.
The produced earthquake catalogs can delineate subsurface fault structures, reveal relationships among earthquakes, and constrain long-term fault behavior \cite{hauksson2012catalog,ross2019ridgecrest,tan2021italy,shelly2023maacama}. To this end, significant efforts have been made to generate more complete, high-resolution earthquake catalogs, including the deployment of dense seismic networks to enhance detection sensitivity \cite{ringler2022network,kohler2020network} and the advancement of source characterization methods \cite{waldhauser2000hypodd,waldhauser2008catalog} to yield high-fidelity results. 

The continuous expansion of seismic monitoring networks, expected to reach new scales through the integration of Distributed Acoustic Sensing (DAS) \cite{zhan2020das}, has led to an exponential surge in the volume of seismic data. This shift necessitates the development of automated monitoring workflows capable of replacing traditional manual processing while archiving comparable or superior performance. 
In particular, the initial earthquake phase picking step, which identifies the first arriving P and S phases, has advanced rapidly. The classic short-term-average / long-term-average (STA/LTA) algorithm \cite{allen1978stalta} can effectively identify impulsive transient signals and remains widely utilized. In recent years, deep-learning-based phase pickers trained on millions of manual labels have improved the sensitivity and precision of extracting phase arrivals from raw seismic waveforms \cite{zhu2019phasenet,ross2018gpd,mousavi2020eqt}. They detect a large number of small-magnitude earthquakes that would otherwise remain hidden, lowering the magnitude of completeness of the resulting catalogs \cite{tan2021italy}. 
The rapidly growing volume of these automatic picks, along with the increasingly large-scale deployment of seismic stations, has posed a crucial challenge to the subsequent phase association step. 
This means that next-generation phase association methods should be resilient to spurious phase picks, capable of resolving complex earthquake sequences, and scalable both computationally to large-scale datasets and spatially from regional to global scales.

Many methods have been proposed for phase association. Among them, the grid-based search methods are most widely used. By first dividing the region of interest into 3D nodes, an earthquake can be declared when enough phase picks are back-projected into a coherent node and hypothetical origin time. Several studies have further improved this strategy. REAL \cite{zhang2019real} reduces the grid-search space of potential events to a smaller volume around the station with the earliest P arrival, assuming that the initiating pick comes from the candidate event. PyOcto \cite{munchmeyer2023pyocto} partitions space-time into a collection of 4D search volumes, repeatedly splits large volumes into smaller ones, discards volumes linked with few picks, and finally identifies the indivisible volumes that contain enough picks to form events. Recently, supervised deep-learning approaches have shown promise for phase association. PhaseLink \cite{ross2019phaselink} uses recurrent neural networks to learn temporal and contextual relationships in sequential picks data. It takes a fixed-length sequence of picks from different stations and predicts whether each pick shares the same event as the first pick. GENIE \cite{mcbrearty2023genie} uses two spatial graphs to represent the station locations and candidate source locations. The graph neural networks are better suited to processing non-Euclidean data structures by allowing the connected nodes to exchange information. 
Unlike the grid-search and supervised methods above, GaMMA \cite{zhu2022gamma} treats association as an unsupervised clustering problem that groups the input phase picks into different clusters. It uses the Gaussian Mixture Model (GMM) to model the probability distribution of each pick and combines the Expectation-Maximization (EM) algorithm to cluster the picks, while simultaneously solving for earthquake location, origin time, and magnitude. However, these methods are mostly limited to local-scale problems and lack generalizability. Any changes to the study region or network geometry requires parameter recalibration, model retraining, or the selection of localized velocity models. Tuning parameters for a specific region is challenging, and the optimized configuration may not transfer to another, or even to the same region with an evolving station geometry. Moreover, the existing phase association methods are coupled either with the repeated location search or with heavy training, so they usually demand substantial computational resources to process large datasets of picks, even though some (e.g., PyOcto) are highly optimized. 

In this study, we propose a new phase association method: the Voronoi tessellation- and Origin-time-based Rapid Associator (VORA). 
We also treat association as an unsupervised clustering problem \cite{zhu2022gamma}, assuming that both P and S arrivals can be identified at the seismic stations surrounding an earthquake. 
At each station, we determine the hypothetical origin times from candidate P and S phase pairs of a common earthquake source. 
Considering that stations near the same source should present coherent hypothetical origin times, the phase association can be simplified as an unsupervised spatio-temporal clustering problem. 
We incorporate the Voronoi diagram, which can be generated from the seismic stations, to establish adaptive neighbor relationships among stations to ensure robust spatial clustering. 
Though the simple clustering works efficiently for most earthquakes, which are separated in space and time from others, dense seismicity scenarios, such as complex aftershock sequences, remain difficult. 
We then estimate the number of candidate earthquakes within each cluster of phase pairs and apply the EM algorithm to assign the phase pairs into the most likely earthquakes. 
Our method is both computationally efficient and governed by fewer hyper-parameters, enabling it to generalize across varying time periods and study regions. Synthetic stress tests and applications to earthquake sequences using both local- and global-scale phase datasets demonstrate that VORA can handle diverse scenarios of seismicity, process large pick volumes, and perform robust from local to global scales. It can thus facilitate the unified seismic monitoring across regions. 

\section{Method}

We treat the earthquake phase association task as a simple spatio-temporal clustering problem, because stations that recorded the same earthquake cluster in space, and their hypothetical earthquake origin times are coherent in time, as shown in Figure~\ref{fig1}.

\subsection{Estimating Earthquake Origin Time}

Each seismic phase arrival time (e.g., $T_P$, $T_S$) at a station is the sum of the earthquake origin time and the phase travel time. Given a known hypocenter and velocity model, the theoretical travel time to a station can be calculated; subtracting this value from the observed arrival time yields the earthquake's origin time ($T_0$). However, in practice, traversing the hypocentral parameter space is computationally demanding, and accounting for complex subsurface velocity structures introduces significant uncertainty. Alternatively,  $T_0$ can be estimated from the Wadati diagram \cite{wadati1933}. Plotting phase arrival-time intervals ($T_S-T_P$) against arrival times ($T_P$) shows a linear relationship where the slope depends on the $V_P/V_S$ ratio and the x-intercept gives the origin time. The advantage of this approach is that a robust approximation of $T_0$ requires  only the measured phase arrivals and a representative $V_P/V_S$ ratio, bypassing the need for an exact velocity structure or hypocenter. Accordingly, we can estimate $T_0$ by combining the $V_P/V_S$ ratio with the mean and difference of $T_P$ and $T_S$  (Eq. \ref{eq1}). 

\begin{equation}
T_0=\frac{(T_P+T_S)}{2}-\frac{(T_S-T_P)}{2}\cdot\frac{(V_P/V_S+1)}{(V_P/V_S-1)}
\label{eq1}
\end{equation}

This approach is robust because the $V_P/V_S$ ratio typically exhibits less regional variation and depth dependence than absolute velocity structures, remaining relatively stable as localized variations cancel out over distance. For instance, utilizing a constant velocity ratio to estimate missing P or S phase arrivals has been common practice before determining precise differential times through waveform cross-correlation \cite{shelly2023maacama}. The approach relying on both $T_P$ and $T_S$ phase arrivals has also been widely applied in single-station detection and location studies \cite{zhang2026kamchatka}, where the distance estimated from the phase arrival-time interval and the calculated back-azimuth are combined to constrain the epicenter. 
This method applies beyond tectonic events to anthropogenic or explosive sources, provided that both P- and S-wave onsets are reported \cite{schweitzer2001hyposat}. 
While S phases are usually more challenging to identify than first-arriving P phases, deep-learning phase picking models have significantly enhanced S-phase detection sensitivity and precision \cite{xi2024phasenet-tf,suarez2025skynet}. 
Recent multi-task deep learning models such as PhaseNet+ \cite{zhu2025phasenet+} have also enabled the direct prediction of origin times from raw single-station waveforms.

\begin{figure}[h]
    \centering
    \includegraphics[width=1.0\linewidth]{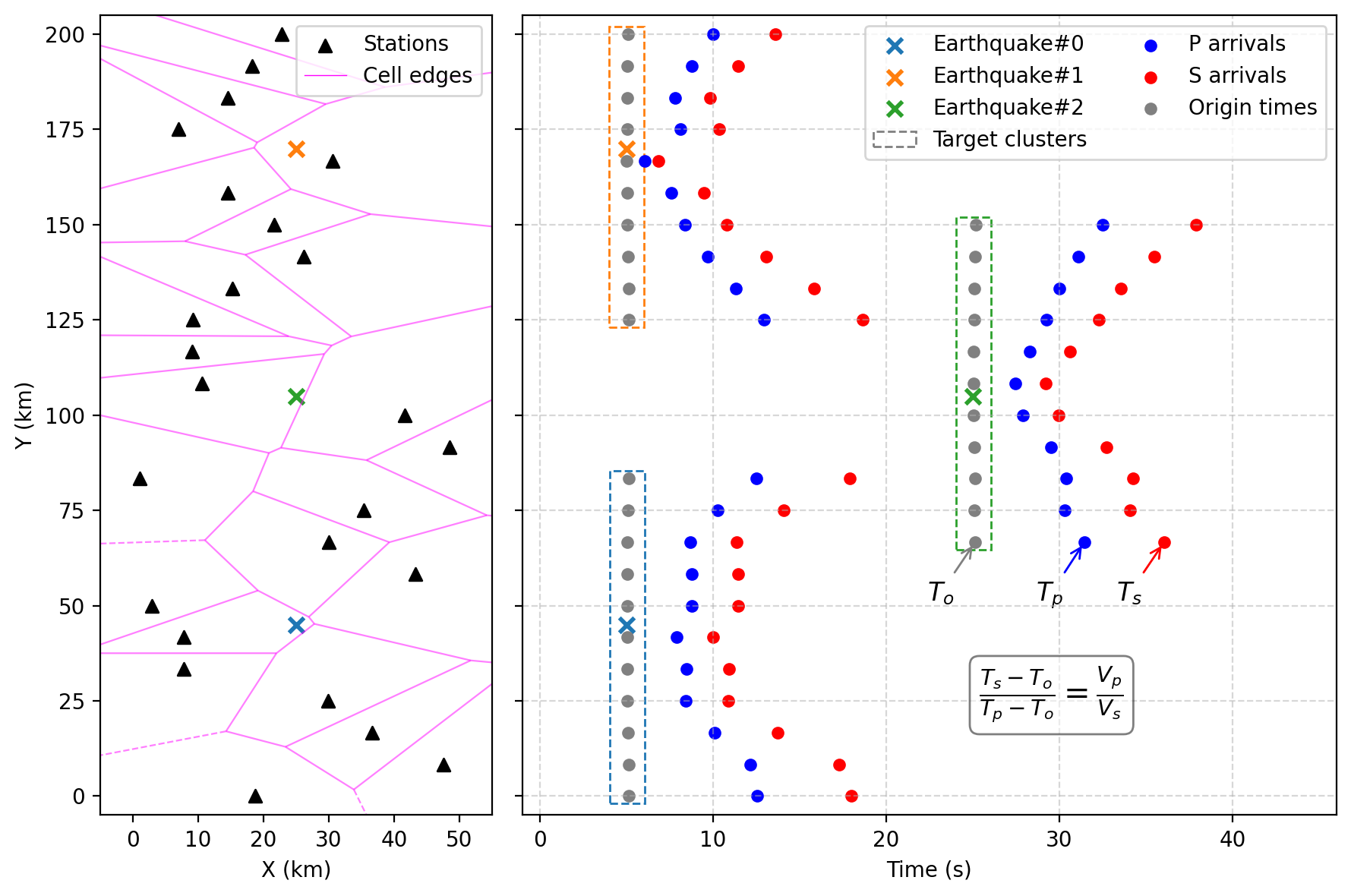} 
    \caption{Schematic view of spatio-temporal clustering of earthquake hypothetical origin times. The left plot shows the Voronoi diagram built using the synthetic stations (black triangles). The origins (time and location) of three synthetic earthquakes are denoted by crosses. Phase arrivals and the estimated earthquake origin times at the nearby stations are shown in the right plot. The hypothetical origin times for each earthquake are close in time and space, forming three target clusters. Phase arrivals associated with each cluster of hypothetical origin times will be eventually grouped together.}
    \label{fig1}
\end{figure}

The hypothetical origin times inferred from the phase pairs at different stations for a common earthquake source tend to cluster in time if phase arrivals are accurately identified and site-specific velocity ratios are appropriately applied (Figure \ref{fig1}). Therefore, a simple temporal clustering algorithm can work well to group them together, which has been successfully demonstrated by \citeA{zhu2025phasenet+} using a histogram method to directly stack the predicted origin times. However, when two or more earthquakes occur close in time (Figure \ref{fig1}), their corresponding hypothetical origin times would be hard to separate. Some previous studies further use the epicentral distances estimated from S-P time differences to determine the hypocenter, an approach known as spatial association \cite{chen2016phasepapy,zhou2022palm}. 
These methods require accurate velocity structures to obtain reliable epicentral distance estimates. 
Another common limitation of these methods is that the detection threshold for candidate earthquakes is strongly influenced by the total number of stations, because more stations bring more false picks into a fixed time window, increasing the false-positive rate. Raising the threshold with the number of stations can reduce false detections but also hinder the detection of small-magnitude earthquakes. 
Despite these limitations, such studies clearly show the potential of using hypothetical origin times to perform earthquake phase association. 

\subsection{Defining Neighbors with Voronoi Tessellation}

Rather than associating the temporally clustered hypothetical origin times into a common hypocenter, we propose a new spatial clustering step that leverages the spatial neighborhood relationships among stations derived from Voronoi tessellation \cite{voronoi1908}. 
A Voronoi diagram partitions a plane (or sphere) into a collection of regions (or cells) based on a set of reference points, such that every point inside a region (or cell) is closer to its reference point than to any other. Similarly, we can treat the seismic stations as the reference points of such a Voronoi diagram. Each station then occupies a Voronoi cell, within which the epicenter of any potential earthquake is closer to this station than to any other stations. Earthquakes that occur at vertices or edges, where two or more Voronoi cells meet, have equal distances to the corresponding stations. We thus define the stations whose Voronoi cells share boundaries as neighboring stations. The distance between neighbors is large for sparse seismic networks and small for the denser ones, 
making the neighborhood adaptive to the spatial density of seismic stations. 
Although the K-Nearest-Neighbors (KNN) algorithm used in previous studies \cite{mcbrearty2023genie} also adapts to inter-station distances, it is less suited to spatially non-uniform networks, because a fixed K could produce asymmetric neighbor relationships, whether the number K is smaller or larger. 
For the same reason, Voronoi tessellation has also been successfully incorporated for adaptive parameterization and regularization of velocity models in seismic tomography \cite{fang2020tomo,mao2023coda}. It has also been used by the real-time earthquake early warning algorithms to constrain evolutionary earthquake locations \cite{satriano2008voro_eew,chen2020voro_loc}.
According to the definition of Voronoi diagram, the resulting station neighborhood is jointly determined by station locations and potential earthquake locations, rather than solely by inter-station distances, resulting in more physically meaningful neighboring relationships when the target is to detect earthquakes.

\begin{figure}[h]
    \centering
    \includegraphics[width=1.0\linewidth]{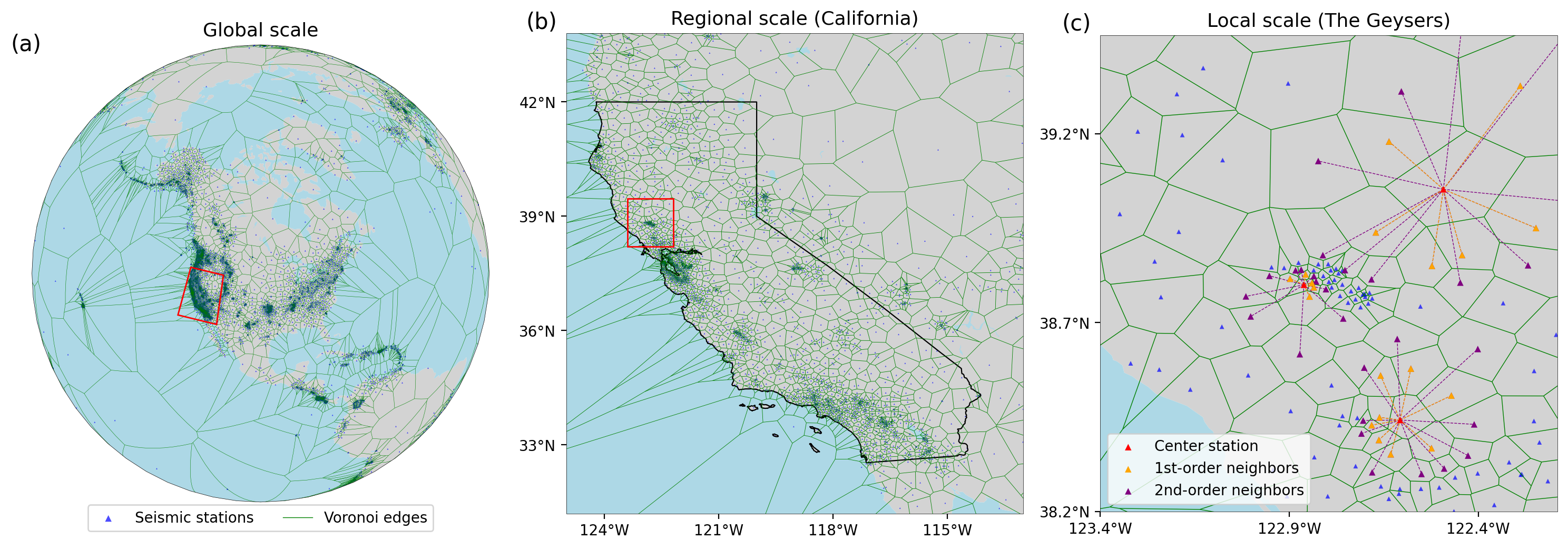} 
    \caption{Multi-scale views of an example Voronoi diagram generated from the locations of globally distributed seismic stations. (a) Global scale, showing a hemisphere centered on North America. Triangles mark the locations of seismic stations, and green lines denote the Voronoi edges that compose Voronoi cell boundaries. (b) A zoomed-in regional view of California. (c) A further zoomed-in local view focusing on The Geysers, illustrating several center stations alongside 1st- and 2nd-order neighbors. Red boxes in (a) and (b) indicate the spatial extents of the subsequent panels.}
    \label{fig2}
\end{figure}

A Voronoi diagram can be generated on a 2D plane or a sphere, so the input is simply the station locations of any local-, regional-, or global-scale seismic network. The diagram can also be easily updated when the network geometry changes, for example, when some stations are added or removed. In Figure \ref{fig2}, we have plotted the Voronoi diagram built based on all the seismic stations operating on the first day of 2025. The global-scale diagram is further zoomed into the regional scale enclosing California and the local scale around The Geysers. It is obvious that the seismic network is highly non-uniform, so the size of Voronoi cells varies significantly among stations. In the local-scale diagram, several stations are chosen as center stations. The stations sharing an edge with a center station are defined as its 1st-order neighboring stations. 
For each center station, we re-construct the Voronoi diagram with its 1st-order neighbors removed, and the new neighbors are defined as the 2nd-order neighboring stations (Figure \ref{fig2}c). To find 3rd-order neighbors, we simply repeat the procedure with 1st- and 2nd-order neighbors removed. 
These higher-order neighboring stations allow reasonable redundancy when certain stations are compromised by noises and fail to capture earthquake signals.
Usually, the 1st- and 2nd-order neighbors are enough to link the stations that have detected the same earthquake.

\subsection{Unsupervised Spatio-Temporal Clustering}


We adopt a modified DBSCAN algorithm \cite{schubert2017dbscan} to perform the unsupervised spatio-temporal clustering. 
The input is a sequence of origin times estimated from Eq.\ref{eq1}, together with the Voronoi neighbor graph. 
Because the true $V_P/V_S$ ratio is not known, we evaluate Eq.\ref{eq1} over a prescribed range of ratios, so each origin time is actually an interval of candidate values rather than a single point. 
For each interval, the neighborhood is the set of intervals at Voronoi-neighbor stations that overlap it in time, which replaces DBSCAN's fixed distance ball. 
We set the minimum number of samples to two, so any interval with at least one overlapping neighbor becomes a core sample; a cluster then grows by a single overlap between adjacent stations and stops only when no further overlap is found. 
Each resulting cluster is one detected earthquake, and the P and S picks of its phase pairs are the associated arrivals (Figure~\ref{fig1}).
Last, we further filter out the clusters with fewer stations than a minimum number of stations, one of VORA's hyper-parameters.
This intuitive and rapid spatio-temporal clustering process effectively discards spurious picks that are coherent in neither time nor space and yields reasonably good associations for routine monitoring when seismicity is at low to moderate levels. 
In denser sequences, however, two earthquakes can occur very close in time and space that this pass merges them (Figure~\ref{fig3}), which the optional step below resolves.

\subsection{Optional Expectation-Maximization-Based Clustering}

\begin{figure}[h]
    \centering
    \includegraphics[width=1.0\linewidth]{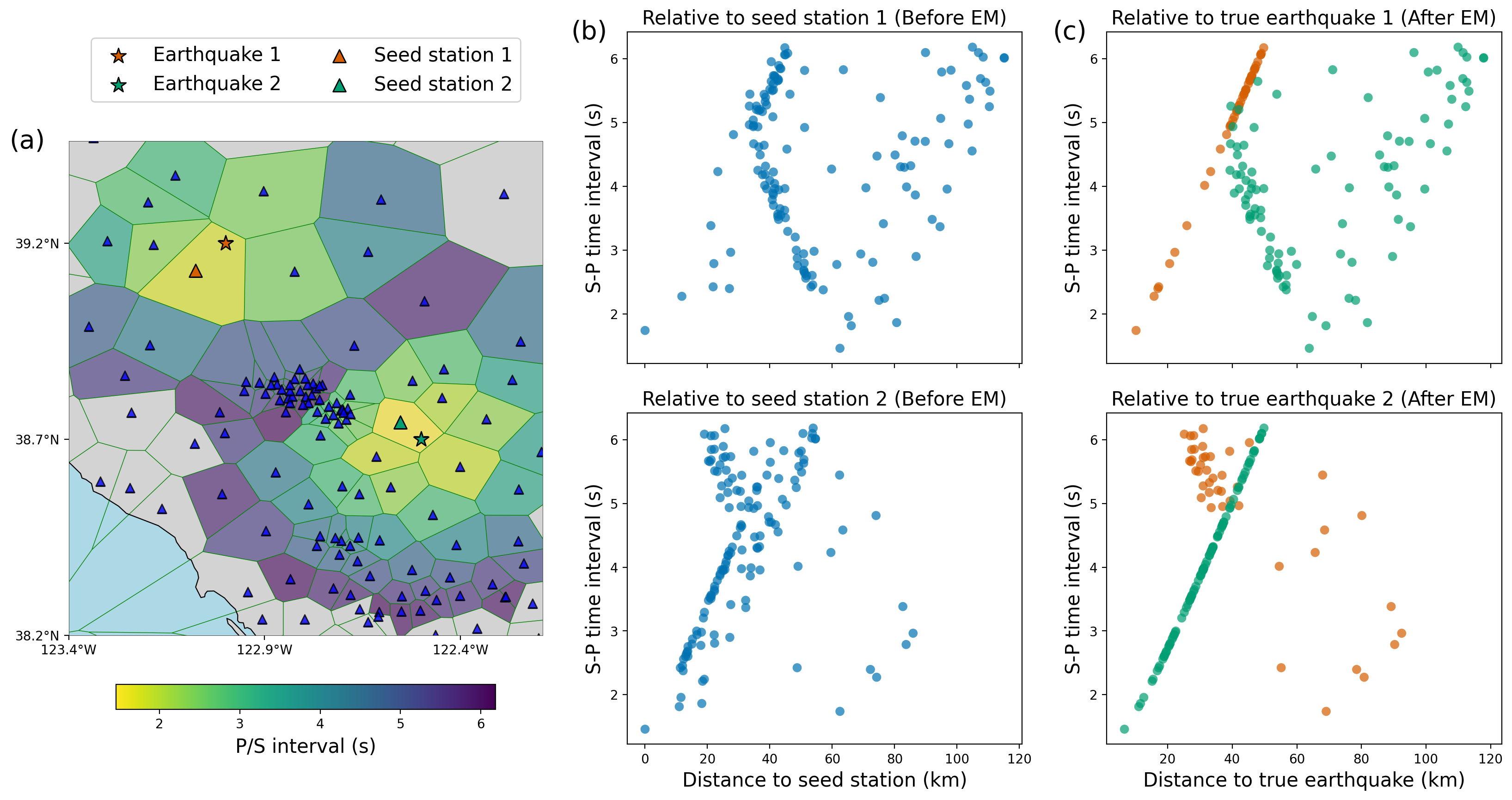} 
    \caption{An example illustrating the optional sub-clustering step for two prescribed earthquakes that happen at the same time, and also close in space based on the EM algorithm. (a) Map showing the locations of two earthquakes (stars) and the same seismic stations (triangles) shown in Figure \ref{fig2}(c). Voronoi cells for the stations detecting the two earthquakes are color-coded by theoretical S-P time intervals, with the two stations (seeds) closest to each earthquake highlighted. (b) S-P time interval versus epicentral distance, assuming the locations of two seed stations as initial epicenters. (c) Results after EM-based clustering. P \& S phase pairs, denoted by S-P time intervals, have been successfully grouped into 2 clusters, corresponding to earthquake 1 \& 2, respectively. Two ground-truth earthquake locations are used as epicenters for better visualization.}
    \label{fig3}
\end{figure}

To address the potential merging issue, we introduce an additional step to perform further clustering based on the Expectation-Maximization (EM) algorithm. 
The key is to first determine how many earthquakes are likely within each cluster. We further exploit the Voronoi diagram and the time intervals of phase pairs. Typically, the travel-time difference of each phase pair increases with distance from the epicenter. If a station's observed time interval is smaller than those from its neighboring stations, the earthquake epicenter is expected to lie within the station's Voronoi cell, or closer to this station than other triggered stations (Figure \ref{fig3}). Within each cluster from the previous step, we detect local minima of the time interval and name the corresponding stations the seed stations. The locations of the seed stations are known, and the candidate hypocenters should be closer to the seed stations. We then incorporate the pseudo slowness ($K$) of the time interval ($\Delta T$) and the earthquake location ($x_e$, $y_e$, $z_e$) into the EM algorithm (Eq. \ref{eq2}). 

\begin{equation}
\Delta T = K\sqrt{(D^2 + z_e^2)}\label{eq2}
\end{equation}
where $\Delta T$ is the observed time interval, $K$ is a slowness constant constrained within a reasonable range, $D$ is the epicentral distance, and $z_e$ is the depth of the source. The range of K can be better constrained since the rough location of the cluster is already determined after the first clustering step.
We then iteratively update the assignments from phase pairs to earthquakes in the E-step and optimize the parameters in the M-step. The initial locations are set to be the locations of seed stations (Figure \ref{fig3}b), and the updates are restricted within the Voronoi cell of seed stations. 
The algorithm iterates until convergence or exceeding the maximum allowed times, after which the sub-clusters of phase pairs are linked to the individual earthquakes (Figure \ref{fig3}c). 
Applying a constant slowness per earthquake is not perfect but proven effective at assigning phase pairs into certain clusters. 
This approach would work better for the region with denser seismic stations because the seed stations will be more adjacent to true epicenters to provide better initial locations. 

\section{Results}

We demonstrate the performance of VORA on 8 synthetic datasets with progressively smaller average time separations between consecutive earthquakes ($\bar{\Delta t}$) and on a real picks dataset for the 2019 Ridgecrest earthquake sequence \cite{zhu2022gamma}. We compare it with two existing methods: GaMMA \cite{zhu2022gamma} and PyOcto \cite{munchmeyer2023pyocto}, in terms of recall, precision, and speed. In addition, we test VORA on a two-week global-scale pick dataset that also covers the 2019 Ridgecrest sequence \cite{ni2025global} to show its spatial scalability. 

\subsection{Synthetic Stress Test}

We follow the stress-testing procedure used by \citeA{ross2019phaselink} to assess VORA under increasing challenging scenarios with progressively shortened $\bar{\Delta t}$. 
For the station locations, we use the local-scale seismic network around The Geysers from Northern California (Figure \ref{fig2} and \ref{fig3}), which consists of 121 stations and shows a relatively non-uniform spatial distribution. We generate 8 synthetic earthquake sequences with hypocenters randomly distributed within a geographic box [-123.39°, -122.19°, 38.21°, 39.46°], spanning a depth range of 0–25 km. Each sequence includes 5000 earthquakes in total, with $\bar{\Delta t}$ gradually reduced across sequences from 64 down to 5 seconds (specifically 64, 32, 16, 12, 10, 8, 6, and 5 s) to simulate increasingly dense seismicity. Each synthetic earthquake is recorded by a number of nearby stations, randomly sampled between 3 and 100. For simplicity and without loss of generality, we use a uniform velocity model with $V_P=6\ km/s$ and $V_P/V_S=1.73$ to calculate the theoretical P and S travel times. Because our method relies on velocity ratio instead of the absolute velocity model, the choice of a constant or 1D velocity profile will not affect our association results. Then, phase arrival times are randomly perturbed by ±0.1 s to represent phase-picking uncertainty, considering the relatively small spatial scale of our study area. We assume that each station always detects both P and S phases from an earthquake, though missing picks are common in real pick datasets. 

\begin{figure}[h]
    \centering
    \includegraphics[width=1.0\linewidth]{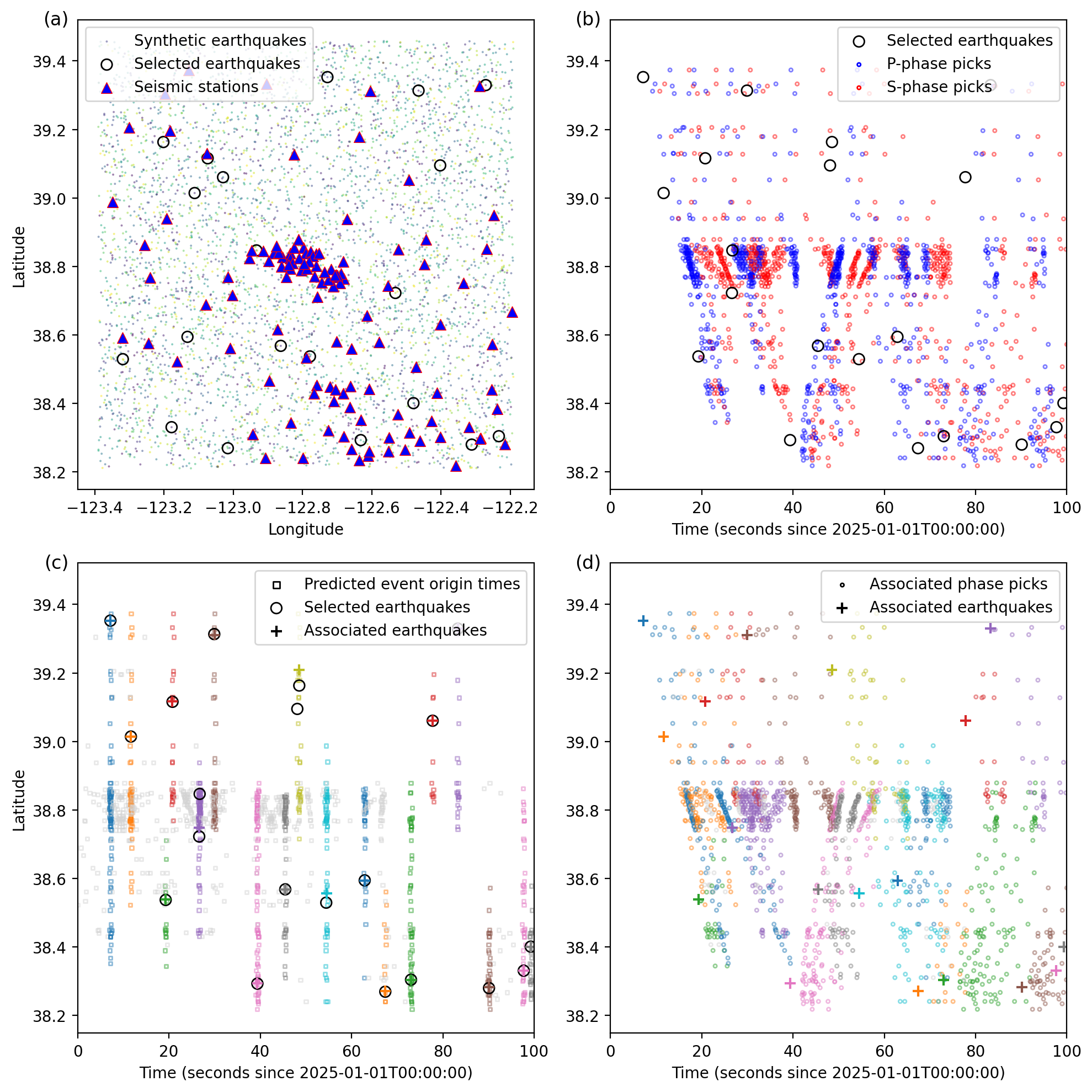} 
    \caption{Synthetic test input and output with a 5-second average time separation between earthquakes ($\bar{\Delta t}$). (a) Map showing the distribution of the 5000 synthetic earthquakes (colored dots) and the same stations shown in Figure \ref{fig2}(c). The open black circles represent 20 earthquakes selected from the first 100-s time window. (b) Synthetic P- and S-phase arrival picks generated by the selected earthquakes, recorded by varying number of nearby stations. (c) Clustering results of the hypothetical origin times (open squares) based on the input synthetic phase picks. Note that a sequence of 10 unique colors are used cyclically for displaying the clusters. The ground-truth and preliminarily located earthquakes are denoted by circles and crosses, respectively. (d) Similar to (b), but phase arrival picks are color-coded based on the association results shown in (c). }
    \label{fig4}
\end{figure}

We first run the VORA association on all 8 synthetic earthquake sequences with neighboring stations considered up to the second order; a $V_P/V_S$ range between [1.63, 1.83]; and at least 3 stations per event. 
Here we show the example input and results for the most challenging scenario of $\bar{\Delta t}=5\ s$, under which the minimum $\Delta t$ is 0 s, and the maximum is 10 s. In the first 100-second time window, 20 synthetic earthquakes are generated within the study region, and their synthetic phase arrivals are close in time and space (Figure \ref{fig4}). 
Based on VORA, 16 of the 20 ground-truth earthquakes have been successfully detected, and the remaining 4 are merged into 2 clusters (Figure \ref{fig4}c). The merged ones can be successfully separated by applying the optional EM-based sub-clustering. However, the quality of the sub-clustering results depends heavily on how precisely the seed stations are identified.

\begin{figure}[h]
    \centering
    \includegraphics[width=1.0\linewidth]{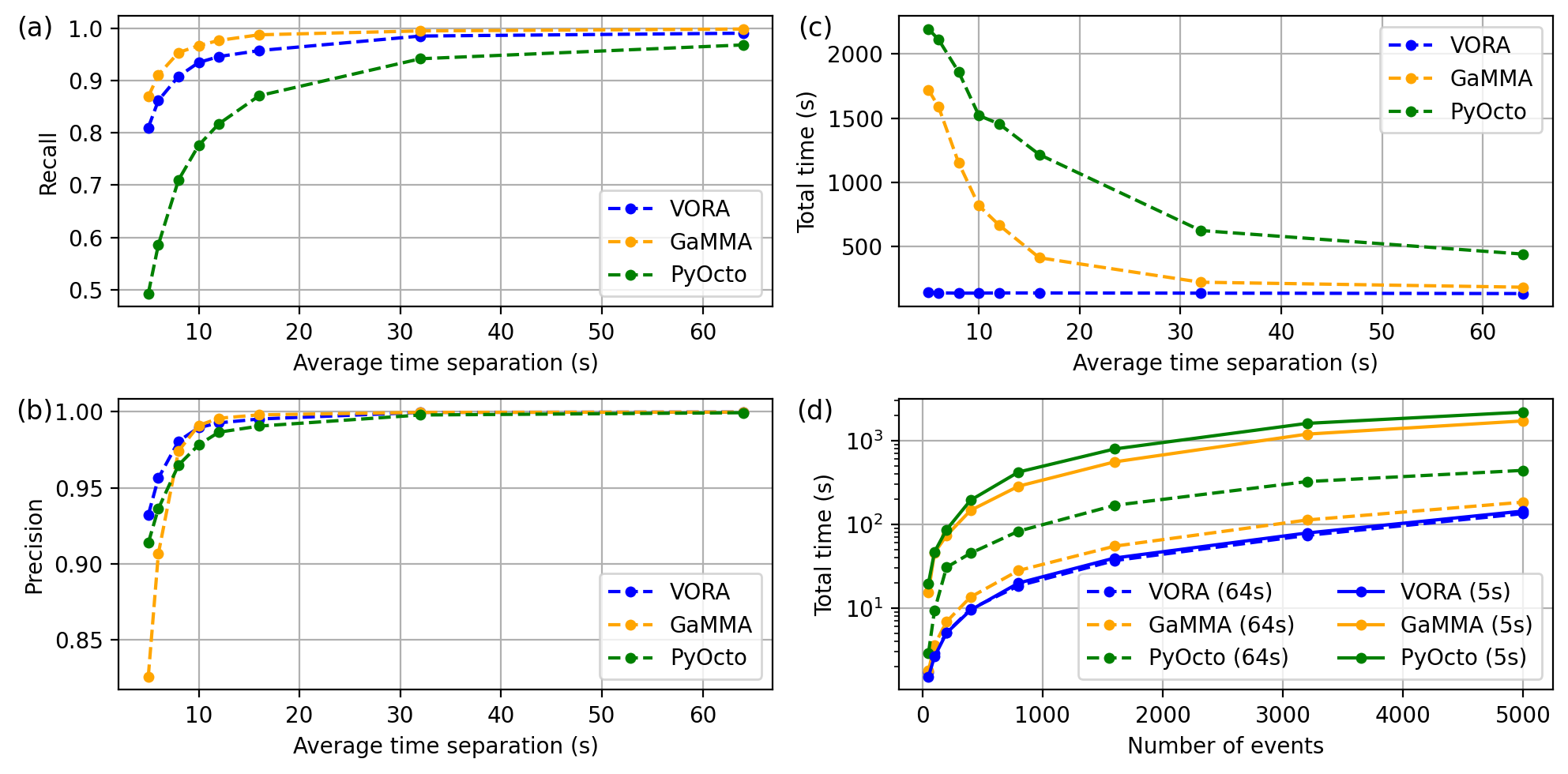} 
    \caption{Stress testing VORA and two other existing methods (GaMMA \& PyOcto) on the 8 synthetic datasets. (a) The changes of recall rate with $\bar{\Delta t}$ (average time separation between earthquakes). (b) The changes of precision rate with $\bar{\Delta t}$. (c) The changes of total run time with $\bar{\Delta t}$ used by each method with single CPU to complete the association.  (d) Association speed comparison across three methods (all with single CPU) for varying number of events. The results for the smallest and largest $\bar{\Delta t}$ are shown with solid and dashed lines, respectively. }
    \label{fig5}
\end{figure}

To benchmark VORA's performance against existing phase association methods, we apply GaMMA \cite{zhu2022gamma} and PyOcto \cite{munchmeyer2023pyocto} to the same 8 synthetic datasets. 
Both methods are given the same velocity model used to generate synthetic phase arrivals. We require at least 3 P and 3 S picks for each associated event. 
We run all 3 methods on a single CPU (Apple M4 Pro, 24 GB RAM) without parallel processing to quantify their runtimes. We obtain 8 earthquake catalogs alongside the associated phase picks for each method. 
We further use the Hungarian algorithm \cite{crouse2016assignment} to find the globally optimal, one-to-one mapping that maximizes pick overlap between true and predicted events. 
For the paired two events, we calculate the recall and precision rates as the ratio of intersecting picks to total picks of the true and predicted events, respectively. The recall rate for unpaired true events and the precision rate for unpaired predicted events would both be 0. A true event is considered a successful recall only if the recall rate of its picks exceeds 0.5. 
Similarly, a predicted event is considered a valid detection when the precision rate exceeds 0.5. For each method and synthetic sequence, the event-level recall rate is defined as the ratio of the number of successfully recalled events to the total number of true events, and event-level precision rate is the ratio of number of valid detections to number of total detected events. 
As shown in Figure \ref{fig5}, both recall and precision rates generally increase with $\Delta t$ for all 3 methods, and when $\Delta t$ is 32 s or larger, the recall and precision rates reach over 0.9 and approximately 1.0, respectively. However, when $\Delta t$ decreases further, the recall rate of PyOcto and the precision rate of GaMMA drop greatly, where VORA maintains recall above 0.8 and precision above 0.9 for the most challenging earthquake sequence (i.e., $\bar{\Delta t}=5\ s$). 
VORA is also faster than the other two methods (Figure \ref{fig5}), which is expected because it is based on unsupervised clustering and decoupled with the repeated locating process. Besides, VORA's runtime does not grow substantially for more challenging synthetic sequences (Figure \ref{fig5}c). For example, at $\bar{\Delta t}=5\ s$, VORA completes the association in less than 2.5 minutes, approximately one order of magnitude faster than GaMMA and PyOcto (Figure \ref{fig5}d). 

\subsection{Test on the 2019 Ridgecrest Earthquake Sequence: Local Scale}

\begin{figure}[h]
    \centering
    \includegraphics[width=1.0\linewidth]{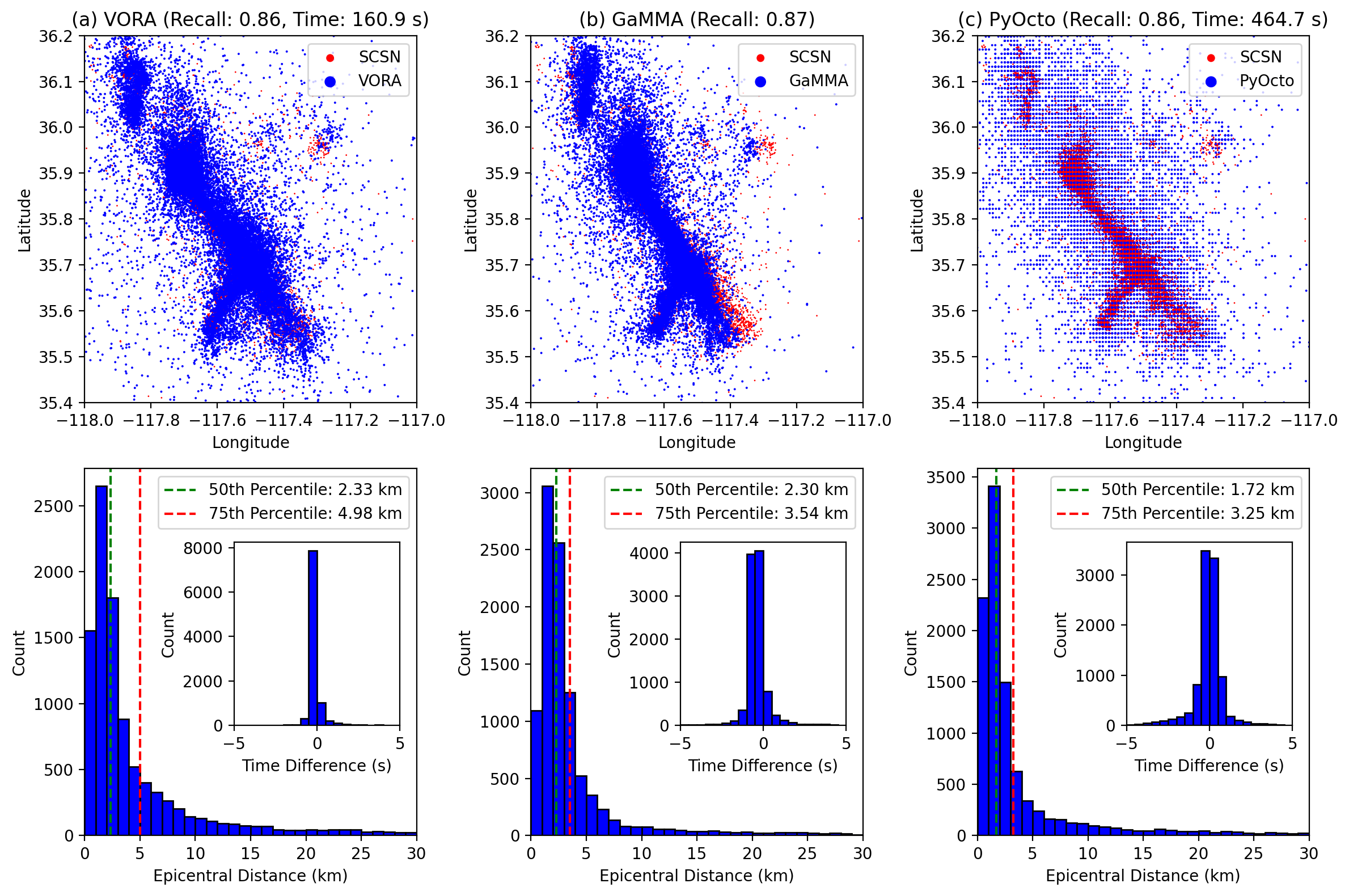} 
    \caption{Association results of VORA and two other existing methods (GaMMA \& PyOcto) on the ML-based phase picks dataset from \citeA{zhu2022gamma} for the 2019 Ridgecrest earthquake sequence. (a) The spatial distribution of associated earthquakes by VORA, and the histograms of origin time difference and epicentral distance compared with the common events from the SCSN catalog. (b) Similar to (a), GaMMA results based on the catalog derived by \citeA{zhu2022gamma}. (c) Similar to (a), PyOcto results derived by this study.}
    \label{fig6}
\end{figure}

The synthetic test allows the controllable experiments and quantitative analysis through
comparison with ground-truth synthetic catalogs and picks. However, it neglects missing picks (false negatives) and noise picks (false positives), which are common in real pick datasets. In this section, we further evaluate VORA on a real machine-learning-based phase pick dataset derived from continuous waveforms of the 2019 Ridgecrest earthquake sequence \cite{zhu2022gamma}. 
To benchmark its performance, we compare it against GaMMA, whose results are obtained directly from \citeA{zhu2022gamma}, and PyOcto, which we apply to the same dataset. In general, VORA's result shows the conjugate faulting architecture, consistent with the first-order structures of relocated earthquake catalogs from previous high-resolution studies \cite{shelly2020ridgecrest,liu2020ridgecrest}. 
We take the SCSN earthquake catalog for the corresponding period as the reference catalog to identify optimally matched event pairs, using maximum thresholds of 5 seconds for origin-time differences and 30 km for epicentral distances. As shown in Figure \ref{fig6}, both VORA and PyOcto achieve a recall rate of 0.86, which is highly comparable to GaMMA's recall of 0.87. For each paired earthquake, we evaluate the epicentral and origin-time discrepancies relative to the reference catalog (Figure \ref{fig6}). While VORA's origin-time residuals are tightly centered on zero, its epicentral distances are systematically larger than those of the other two methods. This trade-off likely stems from VORA's requirement that each station provide both P and S picks, reducing the number of stations available to constrain the hypocentral inversion. Nevertheless, VORA's performance is comparable to these two well-developed methods, without modifying any hyperparameters used in the synthetic test.

\subsection{Test on the 2019 Ridgecrest Earthquake Sequence: Global Scale}

Beyond the speed and robustness shown above, another unique advantage of VORA is that the same framework applies across spatial scales, from local to global scales. In this section, we apply VORA to a subset of the global-scale pick database produced by \citeA{ni2025global}, which used cloud computing and a re-trained PhaseNet \cite{munchmeyer2022whichpicker} model to perform the phase picking. It is worth noting that only the continuous seismic waveforms available from the IRIS/EarthScope, NCEDC, and SCEDC AWS cloud servers are included, meaning that many seismic stations operated by other data centers are absent from this dataset. Processing the entire 20-year dataset is beyond the scope of this method paper. We instead select a representative two-week period, which contains the 2019 Ridgecrest earthquake sequence, and run VORA at the global scale.

\begin{figure}[h]
    \centering
    \includegraphics[width=1\linewidth]{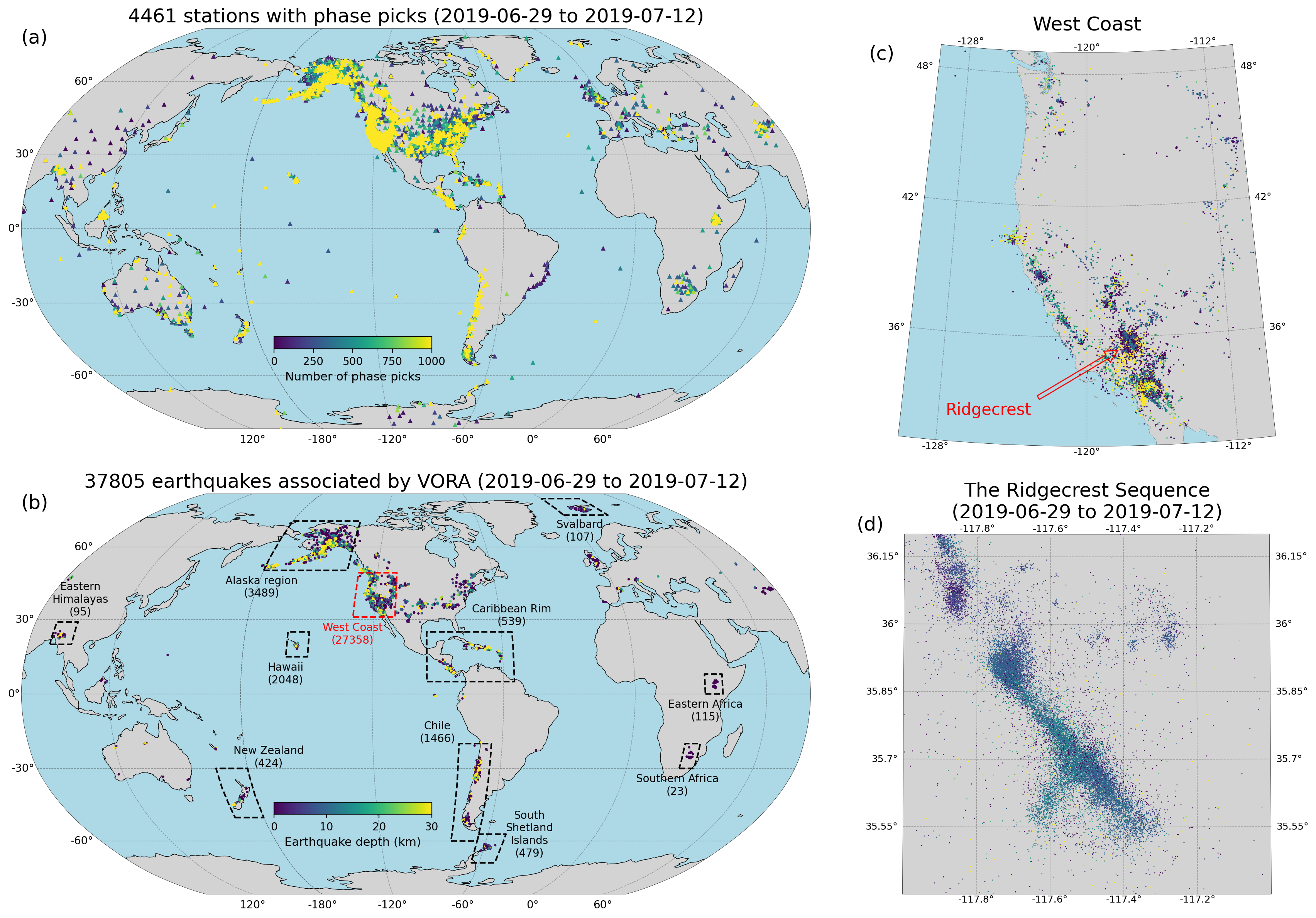} 
    \caption{Association results of VORA on the two-week global-scale ML-based phase picks dataset  \cite{ni2025global} (a) Global distribution of seismic stations, color-coded by the number of picks. Stations without any picks generally lack available waveforms \cite{ni2025global} and are therefore excluded from the association process. (b) Global distribution of associated and preliminarily located earthquakes, color-coded by depth. Dashed boxes outline selected seismically active regions, labeled with their names and earthquake counts; the highly active US West Coast region is highlighted in red. (c) Zoomed-in view of detected seismicity along the US West Coast during the two-week period; the red arrow indicates the Ridgecrest region. (d) Further detailed view of the association results for the 2019 Ridgecrest earthquake sequence within the same geographic bounds as in Figure \ref{fig6}.}
    \label{fig7}
\end{figure}

During the selected period, more than 10 million phase picks are available from 4461 seismic stations (Figure \ref{fig7}a), after removing duplicate picks from the same station recorded on different instruments. We keep the Voronoi diagram dynamically updated at a daily frequency, retaining only stations with phase picks, so that all stations used for building the Voronoi diagram are in operation. We keep the same hyperparameters as in the previous experiments, despite the change from local to global scale. The maximum P-S separation of 20 seconds for phase pairing would thus inherently limit each station's detection range to less than $\sim$1.5 degrees. For this reason, we do not expect to detect regional or teleseismic earthquakes, which are typically several hundred kilometers from the nearest station. On the other hand, the phase picks are also predicted using a deep learning model trained on local earthquake recordings, which lead to degraded phase picking performance when applied to regional or teleseismic earthquakes. 
The dataset also lacks stations across most regions, as many networks do not archive their continuous data with IRIS/EarthScope, and so it is not processed by \citeA{ni2025global}. Even so, it remains the largest-scale pick dataset available to demonstrate our method. 

\begin{figure}[h]
    \centering
    \includegraphics[width=1\linewidth]{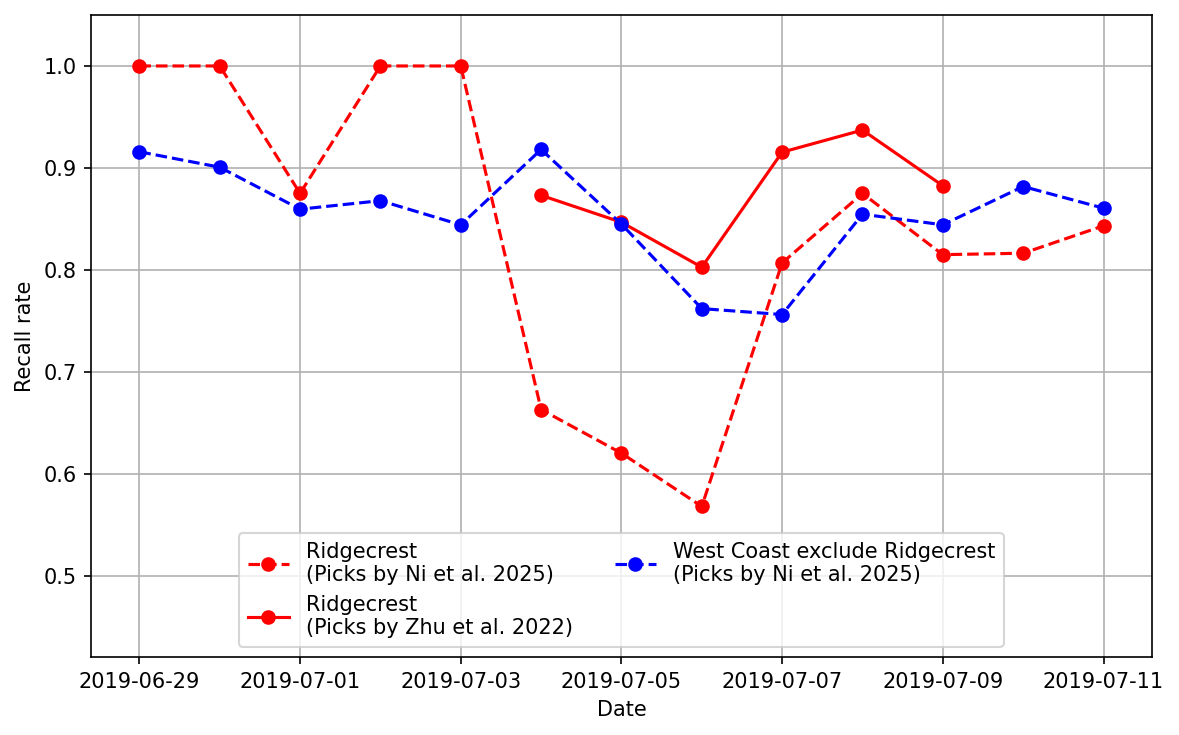} 
    \caption{Daily variations in VORA recall rates across different regions and pick datasets. Red and blue dashed lines represent recall rates using PhaseNet picks from \citeA{ni2025global} for the Ridgecrest and US West Coast (excluding Ridgecrest) regions, respectively. The solid red line denotes the recall rate obtained using PhaseNet picks from \citeA{zhu2022gamma} for the Ridgecrest region. }
    \label{fig8}
\end{figure}

About 37,800 earthquakes are detected for the chosen time period, and their preliminary locations are also derived using the simple constant velocity model ($V_P=6\ km/s$ and $V_P/V_S=1.73$). Most earthquakes are detected within regions that have dense seismic instrumentation and high pick counts (Figure \ref{fig7}a\&b), i.e., regions with both archived data and active tectonics. For example, Alaska and the US West Coast show the highest seismicity, followed by regions such as Hawaii and Chile. 
A zoomed-in view of the 2019 Ridgecrest earthquake sequence shows the same conjugate faulting geometry (Figure \ref{fig7}d), consistent with the results from the local-scale test above. Taking the USGS earthquake catalog as a reference, we calculate the daily recall rates of the VORA association results inside US West Coast (Figure \ref{fig7}c). Before the earthquake sequence, the recall rate could reach 1.0 within the Ridgecrest area and exceed 0.85 across the broader US West Coast region (Figure \ref{fig8}). During the first three days of the sequence, the recall rate within the epicentral area drops to approximately 0.6, then recovers to about 0.8. In comparison, the recall rate outside has not shown significant variations. With the further improvement of pick database \cite{ni2025global} and the integration of appropriate velocity models, higher recall rates are anticipated. Including or excluding stations elsewhere has negligible impact on the US West Coast association results, given that the internal station neighborhood changes exclusively along the boundary. Therefore, VORA can be deployed either independently or jointly for seismic networks operating at local, regional, or national scales, producing highly consistent detection results of local earthquakes within each target region. As illustrated in Figure \ref{fig7}, running VORA directly using a spherical station graph constructed from global stations with available phase picks has provided a unified approach to reveal the seismicity in many different regions, such as New Zealand. This capability offers a promising pathway toward generating earthquake catalogs with highly uniform quality across diverse networks.

\section{Discussion}

The goal of earthquake phase association is precisely grouping together phase arrivals that originate from a common earthquake \cite{ross2019phaselink}. After this step, based on the associated phase arrivals from multiple stations and a proper velocity model, the origin time and location can be further constrained and refined by dedicated location algorithms \cite{waldhauser2000hypodd,schweitzer2001hyposat}.
Earthquake waveforms not only capture phase arrivals and amplitudes but also contain rich information about the source, including origin time \cite{zhu2025phasenet+,zhou2025aipal}, back azimuth, and epicentral distance \cite{ide2026p_onset,zhang2026kamchatka}.
Simple temporal clustering of estimated earthquake origin times has been proposed to facilitate the earthquake phase association procedure \cite{chen2016phasepapy,zhou2022palm,zhu2025phasenet+}. 
However, when the seismic network is large-N or large-scale, simply clustering hypothetical origin times in time has inherent limitations. 
First, the number of false origin time predictions or falsely triggered stations within the fixed time window will increase with the total number of stations. 
A small threshold of the cluster size could produce too many clusters while a larger threshold could miss smaller earthquakes detected by fewer stations.
Second, two or more earthquakes far away from each other and located in regions with completely different geological settings may be included within the same temporal cluster.
Using the velocity model to perform grid-based search of origin locations  \cite{chen2016phasepapy,zhou2022palm} is computationally expensive.
Meanwhile, a single absolute velocity model cannot work appropriately for earthquakes from different regions.
The use of inappropriate absolute velocity models may also produce false associations. 
An additional spatial clustering step that requires the estimated origin times to be spatially coherent can address the above limitations, reducing the threshold issue and bypassing the determination of origin location for each cluster.
Therefore, VORA further leverages the spatial clustering pattern of adjacent stations triggered by the same earthquake to treat it as an unsupervised spatio-temporal clustering problem. 
It is computationally efficient by avoiding the intensive back-projection of individual phase arrivals into potential source origins and locations, and also spatially scalable because the origin-time estimate depends on the $V_P/V_S$ ratio (Eq. \ref{eq1}), which is more stable across regions and depths than the absolute velocity structure.

To ensure robust spatial clustering, we introduce the Voronoi tessellation to build the station graph, which defines the neighboring relationships among stations based on candidate source locations. 
Each station is assigned a specific surrounding region, named Voronoi cell, within which any potential source is closer to that station compared to any other stations \cite{satriano2008voro_eew}.
In other words, given a network of stations, two stations are regarded as neighbors only when their Voronoi cells share a common edge or vertex, where sources are closest to the two stations.
Compared to the KNN-based or distance-based graphs, which considers only the inter-station distance and neglects the adjacency to potential source locations, the station graph based on the Voronoi diagram is more physically meaningful. 
Because the graph is built from whatever stations are operating, it adapts to changing station density and network geometry without retuning, and the same procedure runs unchanged from a local array to a globally distributed set of stations (Figure \ref{fig7}).
We also develop a new way of defining higher-order neighbors of a station by removing its lower-order neighbors, rather than directly taking neighbors of neighbors, since some stations may loss information for reasons like high-level background noise.
Therefore, every neighbor of a station from our station graph is determined based on one Voronoi diagram,
greatly optimizing the information exchange among seismic stations, especially for the earthquake detection problem. 

VORA is governed by only a few parameters, each physically interpretable and each stable against changes in geological region or network geometry. 
The first is the minimum number of stations, corresponding to the minimum allowed cluster size. The second is the highest order of station neighbors, reflecting the extent to which the stations are sensitive to signals and affected by noises. If stations are considered sensitive enough and free from background noise, then 1st-order neighbors are sufficient. 
The third and final one is the estimated origin time and its uncertainty range. It can be directly predicted from continuous raw waveforms using advanced multi-task deep learning models \cite{zhu2025phasenet+}, or computed from one pair of P and S phase arrivals following Eq. \ref{eq1}. Applying a reasonable range of $V_P/V_S$ ratio is an effective method for the successful temporal clustering, because the estimated origin time should have a certain uncertainty range that ideally encloses the true origin time. 
More importantly, VORA's parameters are mainly site-specific, allowing the optimization for individual stations. In comparison, most existing methods rely on a single set of global parameters, 
limiting their adaptability across regions and networks.
Besides, VORA is not limited to using only origin times for time clustering.
Another intriguing possibility is to use the triggers of large amplitude P onsets within very short time windows for earthquake early warning applications. 

\begin{figure}[h]
    \centering
    \includegraphics[width=1\linewidth]{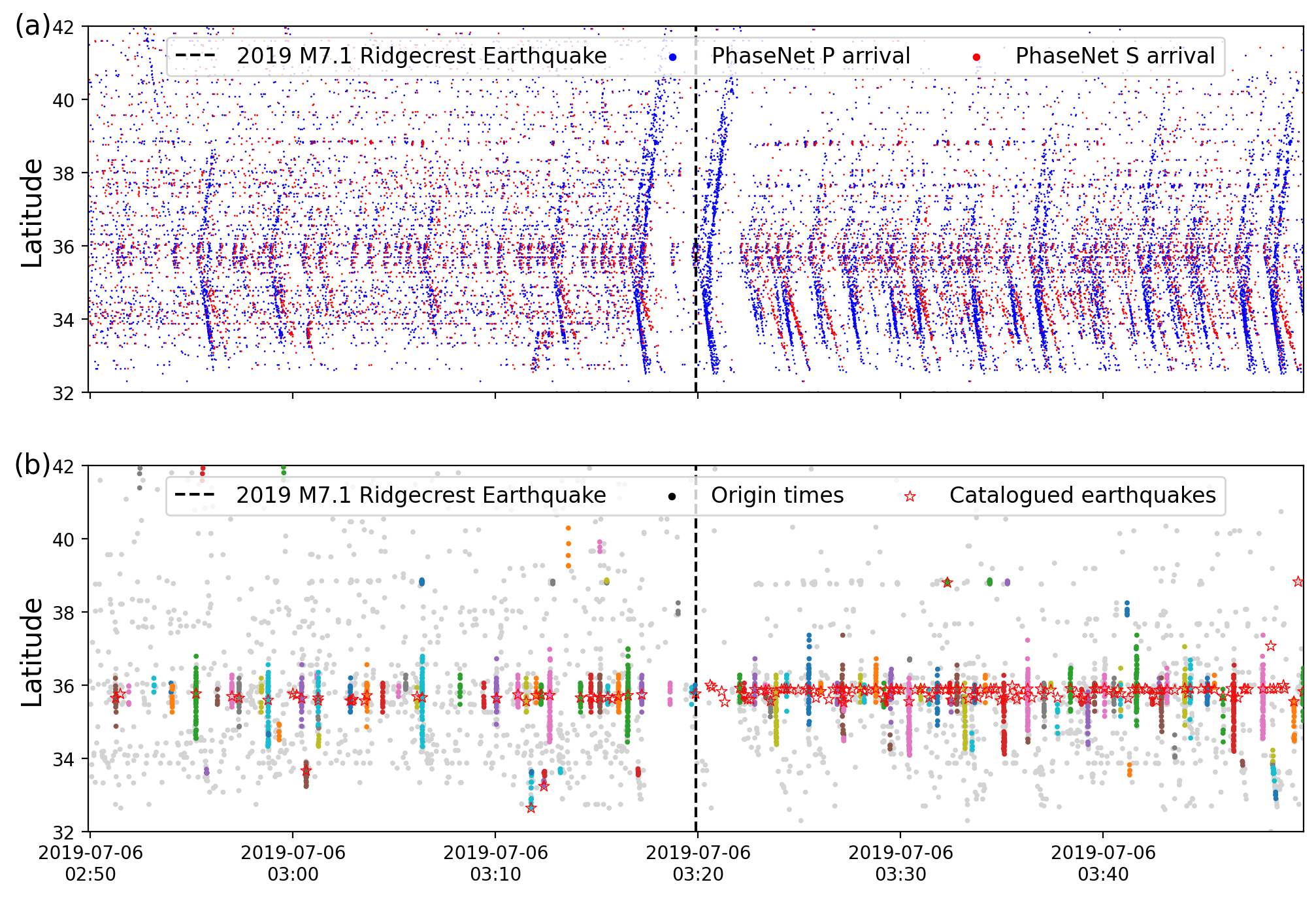} 
    \caption{Example input and output of VORA surrounding the origin of the 2019 Ridgecrest mainshock. (a) All available PhaseNet phase picks \cite{ni2025global} from the stations within the geographic bounds [-130°W, -110°W, 32°N, 43°N], spanning a 1-hour window centered on the mainshock time. (b) VORA association results, represented by clustered hypothetical origin times similar to Figure \ref{fig4}(c) compared against cataloged reference earthquakes for the same region.}
    \label{fig9}
\end{figure}

Compared to the conventional phase picking method like STA/LTA \cite{allen1978stalta}, recent machine learning models have greatly enhanced the detectability of phase arrivals \cite{zhu2019phasenet,mousavi2020eqt}. 
VORA is developed based on the ongoing trend that S phases are getting identified as reliably as P phases \cite{yuan2023ensemble,xi2024phasenet-tf}, or origin times can be directly predicted from raw waveforms \cite{zhu2025phasenet+,ide2026p_onset}. 
However, different models produce phase picks of varying qualities, and usually have downgraded performance when applied to seismic data from other geologic settings \cite{munchmeyer2022whichpicker,zhou2025aipal}. To evaluate the influence of input phase pick quality to VORA, we repeat the daily recall rate analysis for the catalog (Figure \ref{fig6}a) based on the alternative phase pick dataset from \citeA{zhu2022gamma}. While the results also show a similar drop-and-recovery pattern (Figure \ref{fig8}), the recall rate during the Ridgecrest sequence is higher up to about 0.2 than that utilizing the picks from \citeA{ni2025global}. This difference emphasizes that the completeness of phase arrivals can strongly affect the downstream association performance. A qualified phase picking model should be able to detect those high-confidence picks labeled by human analysts, then followed by targeting at identifying more hidden picks. 
A detailed inspection of the 1-hour window around the Ridgecrest mainshock reveals that VORA captures most of the events reported in the routine catalog (Figure \ref{fig9}). However, some earthquakes are still missing, in particular the mainshock itself, mainly due to a deficiency in S phase picks. Although this is an inherent limitation of VORA, it suggests a valuable improvement direction for newly emerging earthquake phase picking models. 
For example, a more generalized phase picking model is still needed for complex overlapping earthquake signals and for regional and teleseismic earthquake waveforms. 

The most difficult case for VORA is several earthquakes clustered in both time and space within the same or adjacent Voronoi cells, which the proposed optional EM-based sub-clustering step separates imperfectly. Denser networks ease this by placing more stations inside each cell. Iteratively removing the best-fitting picks until the remaining picks no longer form a valid event \cite{woollam2020hex} offers another complementary refinement. Shortening the analysis window is a natural extension toward early warning and real-time monitoring, where rapidly resolving overlapping arrivals during vigorous aftershock sequences would be most valuable.

\section{Conclusions}

We have developed a new earthquake phase association method, Voronoi tessellation- and Origin-time-based Rapid Associator (VORA), which back-projects seismic phase pairs to hypothetical origin times and treats association as an unsupervised spatio-temporal clustering problem. We incorporate an optional step to further resolve the merged clusters that are proximate in both space and time by 
re-clustering using the Expectation-Maximization algorithm. Based on the synthetic stress tests  and applications to real picks dataset, we demonstrate VORA's ability to rapidly process large volumes of phase arrivals, perform association from local to global scales, and achieve reasonable results with few hyperparameters to tune. As seismic networks continue to expand and densify, VORA offers a fast training-free path to unified seismicity catalogs across regions and scales supporting earthquake monitoring and research.

\section*{Open Research Section}
The two-week global-scale PhaseNet picks can be accessed from \citeA{ni2025global}. The 6-day PhaseNet picks for the 2019 ridgecrest sequence can be accessed from \citeA{zhu2022gamma}. All codes will be open source on GitHub after the peer review process.

\section*{Conflict of Interest declaration}
The authors declare that there are no conflicts of interest for this manuscript.

\acknowledgments
The present work was inspired by some previous studies \cite{fang2020tomo,mao2023coda,zhu2025phasenet+}.
W. Zhu was supported by the U.S. Department of Energy, Office of Basic Energy Sciences, under Award Number DE-SC0026120.
J. Song was supported by the California Governor’s Office of Emergency Services, under Award Number A251014551.
T. Taira was supported by the United States Geological Survey (Award Number G21AP10152 and G24AP00049) and the Statewide California Earthquake Center (Award Number 25123). 
%
%

\bibliography{references}

@article{hauksson2012catalog,
  title={Waveform relocated earthquake catalog for southern California (1981 to June 2011)},
  author={Hauksson, Egill and Yang, Wenzheng and Shearer, Peter M},
  journal={Bulletin of the Seismological Society of America},
  volume={102},
  number={5},
  pages={2239--2244},
  year={2012},
  publisher={Seismological Society of America}
}

@article{ross2019ridgecrest,
  title={Hierarchical interlocked orthogonal faulting in the 2019 Ridgecrest earthquake sequence},
  author={Ross, Zachary E and Idini, Benjam{\'\i}n and Jia, Zhe and Stephenson, Oliver L and Zhong, Minyan and Wang, Xin and Zhan, Zhongwen and Simons, Mark and Fielding, Eric J and Yun, Sang-Ho and others},
  journal={Science},
  volume={366},
  number={6463},
  pages={346--351},
  year={2019},
  publisher={American Association for the Advancement of Science}
}

@article{tan2021italy,
  title={Machine-learning-based high-resolution earthquake catalog reveals how complex fault structures were activated during the 2016--2017 central Italy sequence},
  author={Tan, Yen Joe and Waldhauser, Felix and Ellsworth, William L and Zhang, Miao and Zhu, Weiqiang and Michele, Maddalena and Chiaraluce, Lauro and Beroza, Gregory C and Segou, Margarita},
  journal={The Seismic Record},
  volume={1},
  number={1},
  pages={11--19},
  year={2021},
  publisher={Seismological Society of America}
}

@article{shelly2023maacama,
  title={Fracture-mesh faulting in the swarm-like 2020 Maacama sequence revealed by high-precision earthquake detection, location, and focal mechanisms},
  author={Shelly, David R and Skoumal, Robert J and Hardebeck, Jeanne L},
  journal={Geophysical Research Letters},
  volume={50},
  number={1},
  pages={e2022GL101233},
  year={2023},
  publisher={Wiley Online Library}
}

@article{ringler2022network,
  title={Achievements and prospects of global broadband seismographic networks after 30 years of continuous geophysical observations},
  author={Ringler, Adam T and Anthony, Robert E and Aster, RC and Ammon, CJ and Arrowsmith, S and Benz, H and Ebeling, C and Frassetto, A and Kim, W-Y and Koelemeijer, Paula and others},
  journal={Reviews of Geophysics},
  volume={60},
  number={3},
  pages={e2021RG000749},
  year={2022},
  publisher={Wiley Online Library}
}

@article{kohler2020network,
  title={A plan for a long-term, automated, broadband seismic monitoring network on the global seafloor},
  author={Kohler, Monica D and Hafner, Katrin and Park, Jeffrey and Irving, Jessica CE and Caplan-Auerbach, Jackie and Collins, John and Berger, Jonathan and Tr{\'e}hu, Anne M and Romanowicz, Barbara and Woodward, Robert L},
  journal={Seismological Research Letters},
  volume={91},
  number={3},
  pages={1343--1355},
  year={2020},
  publisher={Seismological Society of America}
}

@article{waldhauser2000hypodd,
  title={A double-difference earthquake location algorithm: Method and application to the northern Hayward fault, California},
  author={Waldhauser, Felix and Ellsworth, William L},
  journal={Bulletin of the seismological society of America},
  volume={90},
  number={6},
  pages={1353--1368},
  year={2000},
  publisher={Seismological Society of America}
}

@article{waldhauser2008catalog,
  title={Large-scale relocation of two decades of Northern California seismicity using cross-correlation and double-difference methods},
  author={Waldhauser, Felix and Schaff, David P},
  journal={Journal of Geophysical Research: Solid Earth},
  volume={113},
  number={B8},
  year={2008},
  publisher={Wiley Online Library}
}

@article{zhan2020das,
  title={Distributed acoustic sensing turns fiber-optic cables into sensitive seismic antennas},
  author={Zhan, Zhongwen},
  journal={Seismological Research Letters},
  volume={91},
  number={1},
  pages={1--15},
  year={2020},
  publisher={Seismological Society of America}
}

@article{allen1978stalta,
  title={Automatic earthquake recognition and timing from single traces},
  author={Allen, Rex V},
  journal={Bulletin of the seismological society of America},
  volume={68},
  number={5},
  pages={1521--1532},
  year={1978},
  publisher={The Seismological Society of America}
}

@article{zhu2019phasenet,
  title={PhaseNet: a deep-neural-network-based seismic arrival-time picking method},
  author={Zhu, Weiqiang and Beroza, Gregory C},
  journal={Geophysical Journal International},
  volume={216},
  number={1},
  pages={261--273},
  year={2019},
  publisher={Oxford University Press}
}

@article{mousavi2020eqt,
  title={Earthquake transformer—an attentive deep-learning model for simultaneous earthquake detection and phase picking},
  author={Mousavi, S Mostafa and Ellsworth, William L and Zhu, Weiqiang and Chuang, Lindsay Y and Beroza, Gregory C},
  journal={Nature communications},
  volume={11},
  number={1},
  pages={3952},
  year={2020},
  publisher={Nature Publishing Group UK London}
}

@article{ross2018gpd,
  title={Generalized seismic phase detection with deep learning},
  author={Ross, Zachary E and Meier, Men-Andrin and Hauksson, Egill and Heaton, Thomas H},
  journal={Bulletin of the Seismological Society of America},
  volume={108},
  number={5A},
  pages={2894--2901},
  year={2018},
  publisher={Seismological Society of America}
}

@article{zhang2019real,
  title={Rapid earthquake association and location},
  author={Zhang, Miao and Ellsworth, William L and Beroza, Gregory C},
  journal={Seismological Research Letters},
  volume={90},
  number={6},
  pages={2276--2284},
  year={2019},
  publisher={Seismological Society of America}
}

@article{munchmeyer2023pyocto,
  title={PyOcto: A high-throughput seismic phase associator},
  author={M{\"u}nchmeyer, Jannes},
  journal={arXiv preprint arXiv:2310.11157},
  year={2023}
}

@article{ross2019phaselink,
  title={PhaseLink: A deep learning approach to seismic phase association},
  author={Ross, Zachary E and Yue, Yisong and Meier, Men-Andrin and Hauksson, Egill and Heaton, Thomas H},
  journal={Journal of Geophysical Research: Solid Earth},
  volume={124},
  number={1},
  pages={856--869},
  year={2019},
  publisher={Wiley Online Library}
}

@article{mcbrearty2023genie,
  title={Earthquake phase association with graph neural networks},
  author={McBrearty, Ian W and Beroza, Gregory C},
  journal={Bulletin of the Seismological Society of America},
  volume={113},
  number={2},
  pages={524--547},
  year={2023},
  publisher={Seismological Society of America}
}

@article{zhu2022gamma,
  title={Earthquake phase association using a Bayesian Gaussian mixture model},
  author={Zhu, Weiqiang and McBrearty, Ian W and Mousavi, S Mostafa and Ellsworth, William L and Beroza, Gregory C},
  journal={Journal of Geophysical Research: Solid Earth},
  volume={127},
  number={5},
  pages={e2021JB023249},
  year={2022},
  publisher={Wiley Online Library}
}

@article{wadati1933,
  title={On the travel time of earthquake waves.(Part II)},
  author={Wadati, K and Oki, S},
  journal={Journal of the Meteorological Society of Japan. Ser. II},
  volume={11},
  number={1},
  pages={14--28},
  year={1933},
  publisher={Meteorological Society of Japan}
}

@article{zhang2026kamchatka,
  title={Foreshock migration preceding the 2025 Mw 8.8 Kamchatka earthquake: Insights from single-station observations},
  author={Zhang, Ji and Kato, Aitaro and Wang, Wei and Zhang, Miao},
  journal={Geophysical Research Letters},
  volume={53},
  number={7},
  pages={e2025GL120956},
  year={2026},
  publisher={Wiley Online Library}
}

@article{zhu2025phasenet+,
  title={Towards end-to-end earthquake monitoring using a multitask deep learning model},
  author={Zhu, Weiqiang and Song, Junhao and Wang, Haoyu and M{\"u}nchmeyer, Jannes},
  journal={arXiv preprint arXiv:2506.06939},
  year={2025}
}

@article{schweitzer2001hyposat,
  title={HYPOSAT--An enhanced routine to locate seismic events},
  author={Schweitzer, Johannes},
  journal={Pure and Applied Geophysics},
  volume={158},
  number={1},
  pages={277--289},
  year={2001},
  publisher={Springer}
}

@article{chen2016phasepapy,
  title={PhasePApy: A robust pure Python package for automatic identification of seismic phases},
  author={Chen, Chen and Holland, Austin A},
  journal={Seismological Research Letters},
  volume={87},
  number={6},
  pages={1384--1396},
  year={2016},
  publisher={Seismological Society of America}
}

@article{zhou2022palm,
  title={An earthquake detection and location architecture for continuous seismograms: Phase picking, association, location, and matched filter (PALM)},
  author={Zhou, Yijian and Yue, Han and Fang, Lihua and Zhou, Shiyong and Zhao, Li and Ghosh, Abhijit},
  journal={Seismological Society of America},
  volume={93},
  number={1},
  pages={413--425},
  year={2022}
}

@article{voronoi1908,
  title={Nouvelles applications des param{\`e}tres continus {\`a} la th{\'e}orie des formes quadratiques. Deuxi{\`e}me m{\'e}moire. Recherches sur les parall{\'e}llo{\`e}dres primitifs.},
  author={Voronoi, Georges},
  journal={Journal f{\"u}r die reine und angewandte Mathematik (Crelles Journal)},
  volume={1908},
  number={134},
  pages={198--287},
  year={1908},
  publisher={De Gruyter Berlin, New York}
}

@article{fang2020tomo,
  title={Parsimonious seismic tomography with Poisson Voronoi projections: methodology and validation},
  author={Fang, Hongjian and Van Der Hilst, Robert D and de Hoop, Maarten V and Kothari, Konik and Gupta, Sidharth and Dokmani{\'c}, Ivan},
  journal={Seismological Research Letters},
  volume={91},
  number={1},
  pages={343--355},
  year={2020},
  publisher={Seismological Society of America}
}

@article{mao2023coda,
  title={Adaptive coda-wave imaging with voronoi tessellation},
  author={Mao, Shujuan and Ellsworth, William L and Beroza, Gregory C},
  journal={Journal of Geophysical Research: Solid Earth},
  volume={128},
  number={8},
  pages={e2023JB026592},
  year={2023},
  publisher={Wiley Online Library}
}

@article{ni2025global,
  title={A Global-scale Database of Seismic Phases from Cloud-based Picking at Petabyte Scale},
  author={Ni, Yiyu and Denolle, Marine and Thomas, Amanda and Hamilton, Alex and M{\"u}nchmeyer, Jannes and Wang, Yinzhi and Bachelot, Lo{\"\i}c and Trabant, Chad and Mencin, David},
  journal={Seismica},
  volume={4},
  number={2},
  year={2025}
}

@article{shelly2020ridgecrest,
  title={A high-resolution seismic catalog for the initial 2019 Ridgecrest earthquake sequence: Foreshocks, aftershocks, and faulting complexity},
  author={Shelly, David R},
  journal={Seismological Research Letters},
  volume={91},
  number={4},
  pages={1971--1978},
  year={2020},
  publisher={Seismological Society of America}
}

@article{liu2020ridgecrest,
  title={Rapid characterization of the July 2019 Ridgecrest, California, earthquake sequence from raw seismic data using machine-learning phase picker},
  author={Liu, Min and Zhang, Miao and Zhu, Weiqiang and Ellsworth, William L and Li, Hongyi},
  journal={Geophysical Research Letters},
  volume={47},
  number={4},
  pages={e2019GL086189},
  year={2020},
  publisher={Wiley Online Library}
}

@article{xi2024phasenet-tf,
  title={Deep learning for deep earthquakes: insights from OBS observations of the Tonga subduction zone},
  author={Xi, Ziyi and Wei, S Shawn and Zhu, Weiqiang and Beroza, Gregory C and Jie, Yaqi and Saloor, Nooshin},
  journal={Geophysical Journal International},
  volume={238},
  number={2},
  pages={1073--1088},
  year={2024},
  publisher={Oxford University Press}
}

@article{woollam2020hex,
  title={Hex: Hyperbolic event extractor, a seismic phase associator for highly active seismic regions},
  author={Woollam, Jack and Rietbrock, Andreas and Leitloff, Jens and Hinz, Stefan},
  journal={Seismological Society of America},
  volume={91},
  number={5},
  pages={2769--2778},
  year={2020}
}

@article{yuan2023ensemble,
  title={Better together: Ensemble learning for earthquake detection and phase picking},
  author={Yuan, Congcong and Ni, Yiyu and Lin, Youzuo and Denolle, Marine},
  journal={IEEE Transactions on Geoscience and Remote Sensing},
  volume={61},
  pages={1--17},
  year={2023},
  publisher={IEEE}
}

@article{suarez2025skynet,
  title={Picking regional seismic phase arrival times with deep learning},
  author={Suarez, Albert Leonardo Aguilar and Beroza, Gregory},
  journal={Seismica},
  volume={4},
  number={1},
  year={2025}
}

@article{schubert2017dbscan,
  title={DBSCAN revisited, revisited: why and how you should (still) use DBSCAN},
  author={Schubert, Erich and Sander, J{\"o}rg and Ester, Martin and Kriegel, Hans Peter and Xu, Xiaowei},
  journal={ACM transactions on database systems (tods)},
  volume={42},
  number={3},
  pages={1--21},
  year={2017},
  publisher={Acm New York, NY, USA}
}

@article{crouse2016assignment,
  title={On implementing 2D rectangular assignment algorithms},
  author={Crouse, David F},
  journal={IEEE Transactions on Aerospace and Electronic Systems},
  volume={52},
  number={4},
  pages={1679--1696},
  year={2016},
  publisher={IEEE}
}

@article{munchmeyer2022whichpicker,
  title={Which picker fits my data? A quantitative evaluation of deep learning based seismic pickers},
  author={M{\"u}nchmeyer, Jannes and Woollam, Jack and Rietbrock, Andreas and Tilmann, Frederik and Lange, Dietrich and Bornstein, Thomas and Diehl, Tobias and Giunchi, Carlo and Haslinger, Florian and Jozinovi{\'c}, Dario and others},
  journal={Journal of Geophysical Research: Solid Earth},
  volume={127},
  number={1},
  pages={e2021JB023499},
  year={2022},
  publisher={Wiley Online Library}
}

@article{zhou2025aipal,
  title={AI-PAL: Self-supervised AI phase picking via rule-based algorithm for generalized earthquake detection},
  author={Zhou, Yijian and Ding, Hongyang and Ghosh, Abhijit and Ge, Zengxi},
  journal={Journal of Geophysical Research: Solid Earth},
  volume={130},
  number={4},
  pages={e2025JB031294},
  year={2025},
  publisher={Wiley Online Library}
}

@article{ide2026p_onset,
  title={Predictable and unpredictable aspects of earthquakes from P-wave onsets of acceleration seismograms},
  author={Ide, Satoshi and Yoshida, Keisuke},
  journal={Earth, Planets and Space},
  year={2026},
  publisher={Springer}
}

@article{chen2020voro_loc,
  title={Real-time earthquake location based on the Kalman filter formulation},
  author={Chen, Yukuan and Zhang, Haijiang and Eaton, David W},
  journal={Geophysical Research Letters},
  volume={47},
  number={11},
  pages={e2019GL086240},
  year={2020},
  publisher={Wiley Online Library}
}

@article{satriano2008voro_eew,
  title={Real-time evolutionary earthquake location for seismic early warning},
  author={Satriano, Claudio and Lomax, Anthony and Zollo, Aldo},
  journal={Bulletin of the Seismological Society of America},
  volume={98},
  number={3},
  pages={1482--1494},
  year={2008},
  publisher={Seismological Society of America}
}

%
%
%
%
%

\end{document}